\documentclass[a4paper,10pt,oneside,onecolumn,number,preprint,centertitle]{elsarticle}
\usepackage{geometry}
\usepackage{graphics}
\usepackage{subfigure}
\usepackage{amssymb}
\usepackage{amsmath}
\usepackage[dvipsnames]{xcolor}
\usepackage{mathtools}
\usepackage{multicol, multirow}
\usepackage{array}
\usepackage{hyperref}
\usepackage{graphicx}
\usepackage{cuted}
\usepackage{flushend}

\begin{document}
\begin{frontmatter}
\title{The external field effect on the opinion formation based on the majority rule and the $q$-voter models on the complete graph}
\author[usu]{Azhari \corref{cp}}
\ead{azhari@usu.ac.id}
\cortext[cp]{Corresponding author}

\author[brin,pf]{Roni Muslim}
\ead{roni.muslim@brin.go.id}

\affiliation[usu]{
    organization={Department of Physics, Universitas Sumatera Utara},
    city={Medan},
    postcode={20155},
    country={Indonesia}
}

\affiliation[brin]{
    organization={Research Center for Quantum Physics,
    National Research and Innovation Agency (BRIN)},
    city={South Tangerang},
    postcode={15314},
    country={Indonesia}
}

\affiliation[pf]{
    organization={Pesantren Fisika, Hikmah Teknosains Foundation},
    city={Yogyakarta},
    postcode={55281},
    country={Indonesia}
}

\begin{abstract}
We investigate the external field effect on opinion formation based on the majority rule and $q$-voter models on a complete graph. The external field can be considered as the mass media in the social system, with the probability $p$ agents following the mass media opinion. Based on our Monte Carlo simulation, the mass media effect is not strong enough to make the system reach a homogeneous state (complete consensus) with the magnetization $m = 1$ for all values of $p$, indicates that the existence of a usual phase transition for all values of $p$. In the $q$-voter model, the mass media eliminates the usual phase transition at $p \approx 0.21$. We obtain the model's critical point and scaling parameters using the finite-size scaling analysis and obtain that both models have the same scaling parameters. The external field effect decreases both models' relaxation time and the relaxation time following the power-law relation such as $\tau \sim N^{\beta}$, where $N$ is the population size, and $\beta$ depends on the probability $p$. In the majority rule model, $\beta$ follows a linear relation, and in the $q$-voter model, $\beta$ follows a power-law relation.
\end{abstract}

\end{frontmatter}
\section{Introduction}

The general concepts of statistical physics make scholars implement the concepts and rules to understand various socio-political phenomena, and this field is called sociophysics. \cite{galam2012socio,castellano2009statistical}. One of the widely discussed topics in sociophysics is opinion dynamics, a simple rule that describes how opinions are formed and changed due to individual interactions. Simple interaction on the microscopic level can present the macroscopic behavior such as phase transition phenomena. One of the reasons scholars study opinion dynamics models is that it presents statistical physics features such as order-disorder phase transitions, scaling, and universality \cite{cardy1996scaling, muslim2020phase,muslim2021phase,muslim2022opinion}.

Two popular opinion dynamics models in sociophysics are the majority rule and the $q$-voter model. Both models describe an interaction rule of a chosen small group in the population where the agents will adopt the most decisive opinion of the group. From the statistical physics point of view, both models exhibit two different types of phase transition phenomena, namely continuous phase transitions in the majority rule model, and discontinuous in the $q$-voter model for $q > 5$, and always continuous for $q \leq 5 $, regardless of the noise type \cite{nyczka2013anticonformity}. The $q$-voter and the majority models are well-known opinion dynamics models that can represent the dynamics of individual opinions in a community. The $q$-voter model illustrates how a group of agents with a certain number chosen randomly in the population influences another agent to adopt their opinion if they are in a homogeneous state (have the same opinion). The majority rule model developed by Galam describes the rule of the interaction of agents in the population that always follows the majority opinion of the group whatever the opinion form  \cite{galam1986majority,galam1999application,galam2002minority,galam2008sociophysics,chen2005majority,cheon2018dynamical}.

The interesting situation nowadays in the social system is that the existence of the mass media strongly influences changes in individual opinions. It is well known that the mass media, such as television, radio, and Facebook, can change individual opinions. For example, the individual can keep or change their opinion by choosing a presidential candidate based on the information from the mass media. Several works that have considered the external field effect (mass media) on the opinion dynamics model can be seen in Refs. \cite{mazzitello2007effects,candia2008mass,rodriguez2010effects,pabjan2008model,sousa2008effects, crokidakis2012effects,gonzalez2007information,gonzalez2005nonequilibrium,martins2010mass,quattrociocchi2014opinion,pineda2015mass,colaiori2015interplay,sirbu2017opinion,li2020effect,freitas2020imperfect,tiwari2021modeling,civitarese2021external}. It was found that in the absence of mass media, diversity of belief boundaries can increase the capacity of the system to reach a consensus. \cite{pineda2015mass} Other research reveals that as media pressure increases, the system moves from pluralism to global consensus. \cite{colaiori2015interplay}. Another work reported that the role of the media greatly accelerates the spread of opinion. The proportion of supporters of media opinion increases rapidly when the media intervention time reaches a certain value, called the best intervention time.\cite{li2020effect}. \textcolor{black}{A recent study examining the influence of an external field defined on a two-dimensional lattice makes the opinion toward a fixed point. \cite{gimenez2021opinion}}.

The phase transition phenomenon is interesting to study because it can describe a change in social situations, such as the consensus-status quo or agreement-disagreement, which can be observed both in continuous and discontinuous phase transition phenomena. In the case of the phase transition, the existence of the mass media effect can eliminate the usual phase transition \cite{crokidakis2012effects}. In other words, we will not be able to observe the change of two states. At the probability of mass media $p \geq p_t $, all agents at the end having the same opinion (complete consensus) indicates the absence of the usual phase transition. 

This paper investigates the external field effect (mass media) in opinion formation based on the majority rule and the $q$-voter model on the complete graph. Following Ref. \cite{crokidakis2012effects}, we perform the Monte Carlo simulation to investigate the external field effect on the opinion evolution based on the majority rule and the $q$-voter model. Based on our results, in the majority rule model, the effect of mass media is not strong enough to make the opinion reach a homogeneous state. In other words, the usual phase transition exists for all values of $p$. In the $q$-voter model, the mass media effect leads public opinion to be homogeneous (complete consensus) at $p \geq 0.21$, which indicates the existence of the usual phase transition for $p <\sim 0.21$.  We also find the power-law relation between the relaxation time $\tau$ and the population size $N$, $\tau \sim N^{\beta}$, where $\beta$ depends on the probability $p$ and has a different form for both models. For the majority rule model, the parameters $\beta$ and $p$ follow the linear relation  $\beta \sim 0.058\, p$, and for the $q$-voter model, the parameters $\beta$ and $p$ follow the power-law relation $\beta \sim p^{0.023}$.

\section{Model and Method}
We investigate the external field effect on opinion formation based on the majority rule and the $q$-voter models on the complete graph. \textcolor{black}{The complete graph is a network structure where links or edges completely connect all graph nodes. In social network science, the network nodes can represent attributes of social systems such as agents or individuals, and the links or edges can represent social connections such as friendship \cite{lewis2011network}. Therefore in the complete graph, all agents are directly connected and can interact with each other with the same probability. This concept is similar to the mean-field in statistical physics, where the system state can be described completely by only one parameter, such as the density of spin-up of the system \cite{nyczka2013anticonformity}}. \textcolor{black}{The complete graph is less realistic than other networks, such as a scale-free network for describing real social networks. However, we can explore some basic physical features, such as the critical points that separate two particular states, phase transition, scaling, and universality class of the system. In addition, in a particular situation such as a large-closed community, with the development of information technology, each individual can be connected to form a complete social connection similar to the complete graph character \cite{trestian2017towards}. From the social context point of view, the external field can be considered the mass media influence. It is well-known that the mass media significantly influences the individual opinion in the population, i.e., individuals keep or change their opinions due to the effects of mass media \cite{pinkleton1998relationships}.}

In the original majority rule model (the model without external field), a small group consisting of a few agents in the population are chosen randomly, where agents will follow the majority opinion of the group, whatever the opinion form \cite{galam1999application}. {\color{black}As in the previous section, we will consider a scenario where the majority rule and the $q$-voter models will always undergo a continuous phase transition when a 'noise parameter' is involved in the model, namely when a group of the agent consists of maximum six agents \cite{nyczka2012phase}. Therefore, this paper considers a specific case where the small group consists of six agents}. The microscopic interaction of the agents based on the majority rule model is described as follows: Six agents in the population are chosen randomly, and two agents will follow the majority opinion of the four agents. {\color{black} Otherwise (in a tie situation), with the probability $p$, agents follow the mass media opinion. The illustration of the model is exhibited in Eq.~\eqref{eq1}.
\begin{align}\label{eq1}
    & \text{(a)} \quad \uparrow \uparrow \downarrow \uparrow \cdots \downarrow \downarrow \quad  \rightarrow \quad     \uparrow \uparrow \downarrow \uparrow \cdots \uparrow \uparrow \nonumber \\
    & \text{(b)} \quad \downarrow \downarrow \downarrow \uparrow \cdots \downarrow \uparrow \quad  \rightarrow \quad     \downarrow \downarrow \downarrow \uparrow \cdots \downarrow \downarrow \\
    & \text{(c)} \quad \downarrow \downarrow \uparrow  \uparrow \cdots \downarrow   \downarrow  \quad  \xrightarrow[]{p} \quad     \downarrow \downarrow \uparrow  \uparrow \cdots \uparrow \uparrow \quad (\text{the external field effect}). \nonumber
\end{align}
The symbols $\uparrow$ and $\downarrow$ represent the agent's opinion with opinion 'up' and 'down', respectively. Eq.~\eqref{eq1} only displays $3$ of $2^6 = 64$ possible agent combinations as an illustration, because each agent has two possible opinions represented by Ising spin $S=\pm 1$. The first column of Eq.~\eqref{eq1} is the initial condition, and the second one is the final condition (after the update). Row (c) in Eq.~\eqref{eq1} shows how the external field (mass media) works to make two agents adopt an 'up' opinion when the four agents are in a tie situation (no majority opinion). The tie situation is so-called as a part of the absence of social validation in the social-psychology context \cite{shrauger1968social}.}

In the original $q$-voter model, $q$-sized agent ($S_q$) formed a committee, chosen randomly in the population to influence another randomly chosen agent called a voter ($S_{q+1}$). {\color{black} The voter will adopt the committee's opinion if they are in a homogeneous state (all agents have the same opinion). However, if the committee's opinion is inhomogeneous, the voter still changes its opinion with probability $\epsilon$ \cite{castellano2009nonlinear}. In this model, we still follow the original model as described in Ref.\cite{castellano2009nonlinear}, but we set $\epsilon = 0$ as in Refs.\cite{civitarese2021external,nyczka2012phase}. The illustration of the model is exhibited in Eq.~\eqref{eq2}
\begin{align}\label{eq2}
    & \text{(a)} \quad  \uparrow \uparrow \uparrow \uparrow \uparrow \cdots \downarrow \quad  \rightarrow \quad      \uparrow \uparrow \uparrow \uparrow \uparrow \cdots \uparrow \nonumber \\
    & \text{(b)} \quad  \downarrow \downarrow \downarrow \downarrow \downarrow  \cdots \uparrow \quad  \rightarrow \quad \downarrow \downarrow \downarrow \downarrow \downarrow  \cdots \downarrow  \\
    & \text{(c)} \quad  \downarrow \downarrow \downarrow  \uparrow  \uparrow \cdots \downarrow \quad  \xrightarrow[]{p} \quad    \downarrow \downarrow \downarrow  \uparrow  \uparrow \cdots \uparrow \quad (\text{the external field effect}). \nonumber
\end{align}
Equation~\eqref{eq2} only displays $3$ of $64$ possible agent combinations of the model as an illustration, where the row (c) displays how the external field works, namely when the committee ($q = 5$) is an inhomogeneous state. With probability $p$, the voter follows the mass media opinion.}

In the illustrations of the model in Eqs.~\eqref{eq1} and $\eqref{eq2}$, we consider the mass media influences the agent(s) to follow its opinion, namely adopts the 'up' ($S_i = +1$) opinion. Therefore, the external field in the model leads to the system reaching a homogeneous state with the magnetization $m = 1$. In the real social context, the scenario of both models can represent a situation of four or five individuals as the internal influence (for example, by family, friends, and colleagues) is stronger than the external influence by the mass media. Therefore the probability of agents changing their opinion is $1$, as exhibited in scenarios (a) and (b) of Eqs.~\eqref{eq1} and \eqref{eq2}. Moreover, we only set the probability $p = \{0,1\}$ when the external field works, where $0$ represents without external field effect and $1$ represents the strongest external field effect.

The evolution of the opinion per site can be analyzed using the order parameter $m$ (magnetization), which can be expressed as:
\begin{equation}\label{eq:magnetization}
    m = \sum_{i =1}^{N} \dfrac{ S_i}{N},
\end{equation}
\textcolor{black}{where $N$ is a total agent (total population) and $S$ is the Ising spin that represents two possible opinions of each agent. The evolution of the order parameter $m$ will be measured in the Monte Carlo step, where every step, the magnetization $m$ increases by $1/N$, decreases by $-1/N$, or remain constant}. We will check that if there is a usual phase transition in this model, then for a small initial density of spin-up $k = 0.01$ or the magnetization value is close to $-1$. We will find that the system never reaches a homogeneous state with $m = +1$. This procedure is similar to the previous work, where the mass media effect is applied to the two-dimensional Sznajd model and observed the usual phase transition for a certain mass media probability threshold $p_t$ \cite{crokidakis2012effects}. In other word the mass media effect eliminates the usual phase transition for $p \geq p_t$, namely for a small initial density of spin-up $k = 0.01$, the system reaches a homogeneous state with $m = +1$ for $p >\sim p_t$. Intuitively, for $p = 0$ (the case without the external field effect), the phase transition occurred at the spin up density $k = 0.5$, namely we will find that the system reaches a homogeneous state with $m = -1$ for $k < 0.5$ and $m =  +1$ for $ k > 0.5$ for a large system \cite{stauffer2000generalization}.

{\color{black}We consider the finite-size scaling relation of the order parameter $m$ to find the critical point $k_c$, the scaling parameters of both models. The finite-size scaling relation can be expressed as \cite{sousa2008effects}:

\begin{align}
    & m(k,N)  = N^{-\delta_1}  \phi((k-k_c)\,N^{-\delta_2}), \label{eq4} \\
    & k_c(N) = k_c + \delta_1 N^{-\delta_2}, \label{eq5}
\end{align}
where $\delta_1$ and $\delta_2$ are the scaling parameters that make the best collapse of all data. The universality class of the model can be defined using the scaling parameters of the model.}

Another interesting parameter in the Monte Carlo simulation is the relaxation time $\tau$, namely the time needed for all agents to reach a homogeneous state (all agents, in the end, have the same opinion). We analyze the relation between the relaxation time $\tau$ to the population size $N$ of both models.

\section{Result and Discussion}
\begin{figure}[t]
    \centering
    \includegraphics[width = 7 cm]{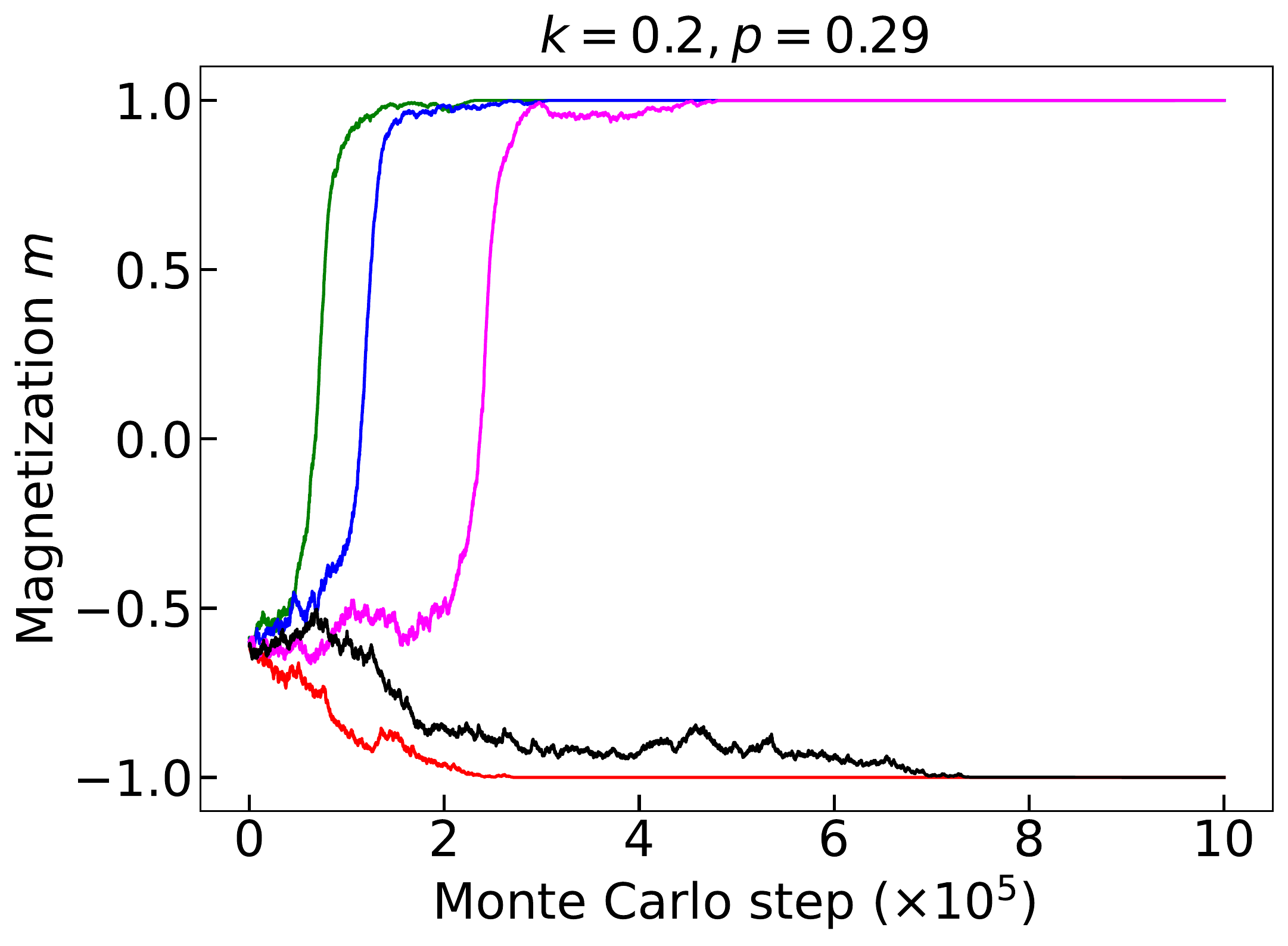}
    \caption{Evolution of the magnetization $m$ under the external field effect based on the majority rule model on the complete graph. As seen, for the initial of spin-up density  $k = 0.2$ and the probability of the external field effect $p = 0.29$, there are several samples reach a homogeneous state with $m = 1$ indicates that the critical point is no longer at $k =k_c = 0.5$ as at the case without the external field effect ($p = 0$), but depends on the probability $p$. The system size $N = 5000$.}
    \label{fig:mcs}
\end{figure}

{\color{black}In order to analyze the phase transition of both models, we first analyze the evolution (in the Monte Carlo step) of the order parameter $m$ defined in Eq.~\eqref{eq:magnetization} under the influence of the external field $p$}. In this part, we only show the majority rule model case for typical values of the spin-up density $k = 0.2$ and the probability of the external field $p = 0.29 $. As seen in Fig.~\ref{fig:mcs}, there are several samples that show the system reach a homogeneous state with $m = 1$ and $m = -1$, indicating the critical point $k_c$ is no longer at  the density of spin-up $k = k_c = 0.5$, such as in the case without the external field effect ($p = 0$), but shifted to another $k$ (somewhere). Therefore, if the phase transition occurs in this model, the critical point depends on the probability $p$.

\begin{figure}[t]
    \centering
    \includegraphics[width = 10 cm]{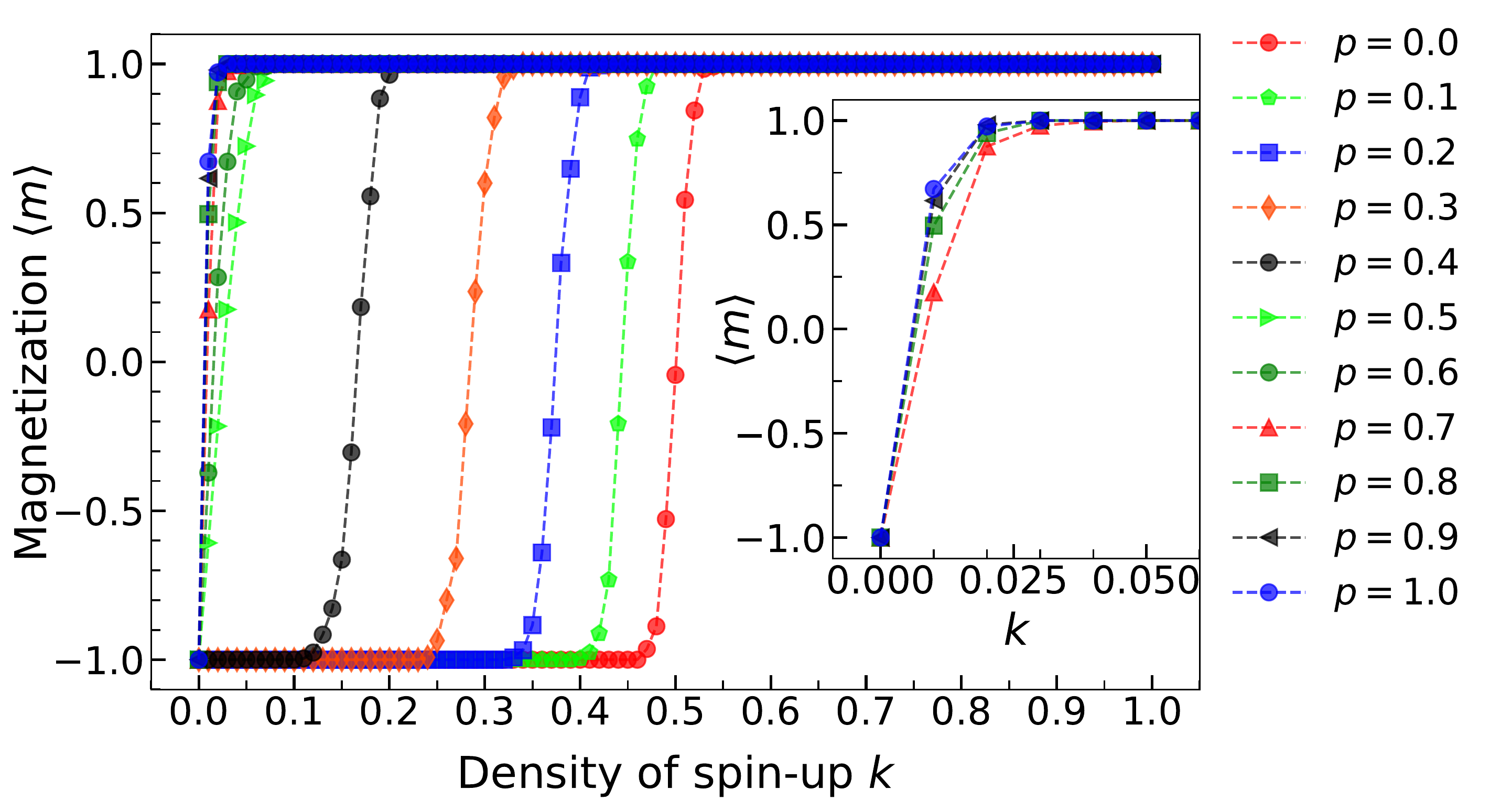}
    \caption{The numerical result of the average of magnetization $m$ versus density of spin-up $k$ for typical values of the probability of the mass media effect $p$ is based on the majority rule model on the complete graph. As seen, the phase transition is observed for all values of $p$, namely for a small initial density of spin-up $k = 0.01$, the system never reaches a homogeneous state with $m = 1.0$ even for maximum $p = 1$. The inset graph exhibits in more detail for $p = 0.7, 0.8, 0.9$ and $p = 1.0$. One can see that there is data that reaches a homogeneous state with $m = 1$ at $k = 0.01$. Each data point averages $500$ samples, and the population size $N = 10000$. }
    \label{fig:fraction_m}
\end{figure}

To see more clearly the external field effect in this model, we variate the density of spin-up $k$ for each typical probability $p$ to obtain the average of $m$. We will analyze that if there is a phase transition in this model, for a small density of spin-up $k$, we will obtain the magnetization $m$ is less than $1$  for each probability $p$. We consider for a large population $N = 10000$, and each data point averages $500$ samples. In the case of the majority rule model, the numerical result of the magnetization $m$ is exhibited in Fig.~\ref{fig:fraction_m}. {\color{black} One can see for the small of initial spin-up density $k = 0.01$, there is no data for all values of $p$ which show the system reaches an homogeneous state with $m =  1$, even for maximum of the external field effect $p = 1$.  This condition means that the presence of the external field in this model is not strong enough to eliminate the phase transition. In other words, the usual phase transition exists for all values of $p$. This phenomenon is caused by the update rule of the majority rule model as shown in Eq.~\eqref{eq1}, relatively not solid compared to the $q$-voter model (will be discussed in the next part) such as shown in Eq.~\eqref{eq2}. Therefore the system is relatively challenging to reach a homogeneous state with  $m = 1$, even for $p = 1$}. From the $64$ possible agent combinations, there are $24$ of agent combinations that will be affected by the external field $p$; that is when the first four agents in Eq.~\eqref{eq1} are in a tied state. There are $40$ other agent combinations that will make the system reach a homogeneous state with $m = + 1$ and $m = -1$ with the same probability. The inset graph of Fig.~\ref{fig:fraction_m} displays in more detail of the data for the range $k = 0.00 - 0.05$ and find the average of the magnetization $m$ for typical values of $p$ at $k = 0.01$  are $ m \approx 0.176 \, (p = 0.7), m \approx 0.496\, (p = 0.8), m \approx 0.616\, (p = 0.9), m \approx 0.672\, (p = 1.0)$. Even for maximum $p = 1.0$, the difference of the magnetization value with the homogeneous state ($m = 1$) is very large, that is $|\Delta m| = 0.328$. {\color{black} However, the exact threshold value of $p$, which eliminates the usual phase transition, is relatively difficult to determine numerically.}

\begin{figure}[t!]
    \centering
    \includegraphics[width = 0.9\textwidth]{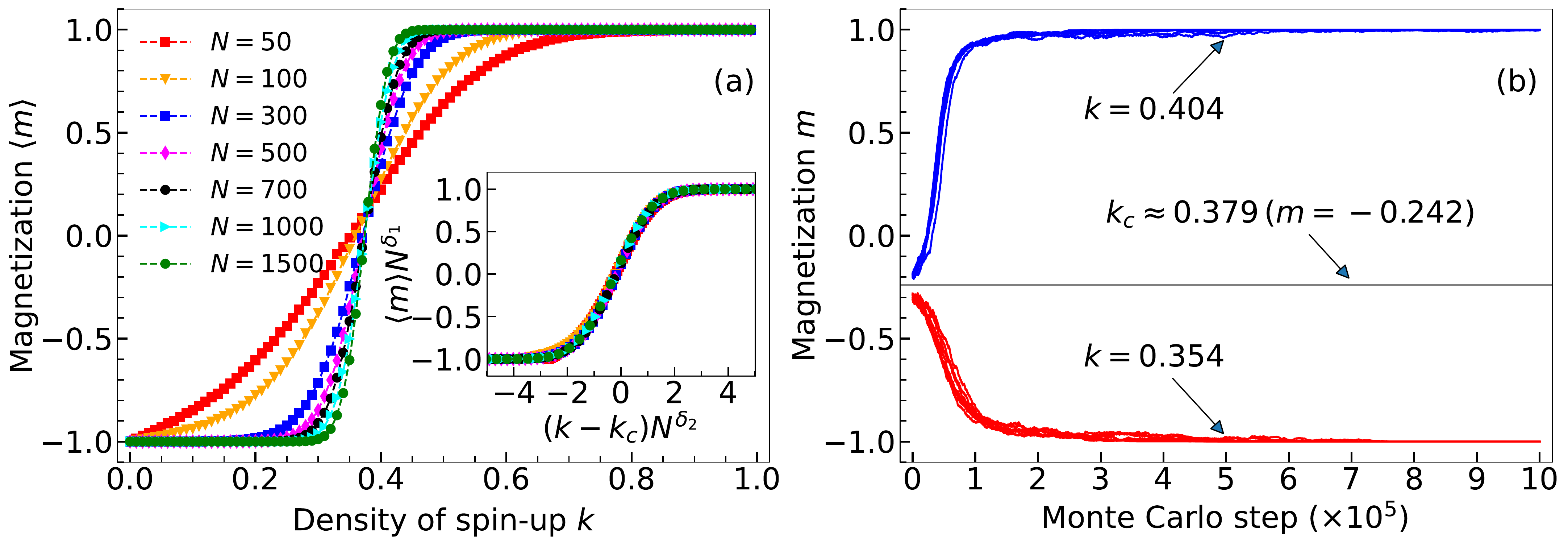}
    \caption{{\color{black}(a) Numerical result of the average magnetization $m$ versus the density spin-up $k$ of the majority rule model for the probability mass media $p = 0.2$ and typical of the population size $N$. The critical point $k_c \approx 0.379$ is obtained from the crossing of lines $m$. The inset graph displays the scaling plot of $m$. Using finite-size scaling analysis, the parameters scaling $\delta_1$ and $\delta_2$ that make the best collapse of all data are $\delta_1 \approx 0, \delta_2 \approx 0.5$. (b) The evolution of the magnetization $m$ for the initial spin-up value $k$ near the critical point $d_c$. One can see for $k = 0.345 < k_c $ and $(k = 0.404 > k_c)$ all samples reach a homogeneous state with magnetization $m = -1$ (red color) and $m = +1$ (blue color).}}
    \label{fig:variate_N}
\end{figure}

{\color{black}We use the finite-size scaling relation in Eqs.~\eqref{eq4}-\eqref{eq5} to obtain the critical point and scaling parameters of this model. We vary the value of $N$ for a typical small probability value of $p = 0.2$. At this point, we will observe the intersection point of lines of the magnetization $m$ indicating the critical point of the system. We find that the critical point of the system is occurred at the spin-up density $k = k_c \approx 0.379$ (see Fig. \ref{fig:variate_N} (a)). This critical point separates two homogeneous states with $m = -1\,(k < k_c)$ and $m =  1\,(k > k_c)$. The scaling parameters that make the best collapse of all data are $\delta_1 \approx 0$ and $\delta_2 \approx 0.5$. These scaling parameters are universal, in other words, the scaling parameters $\delta_1 \approx 0$ and $\delta_2 \approx 0.5$ are valid for all values of $N$ and $p$.  We check the evolution of the magnetization $m$ by giving the initial spin-up density $k$ near the critical point $k_c \approx 0.379$ for a large population $N = 10000$, as exhibited in Fig.~\ref{fig:variate_N}~(b). One can see that for $k < k_c \, (k > k_c) $ all samples reach a homogeneous state with magnetization $m = - 1 (m = 1)$.}

\begin{figure}[t!]
    \centering
    \includegraphics[width = 10 cm]{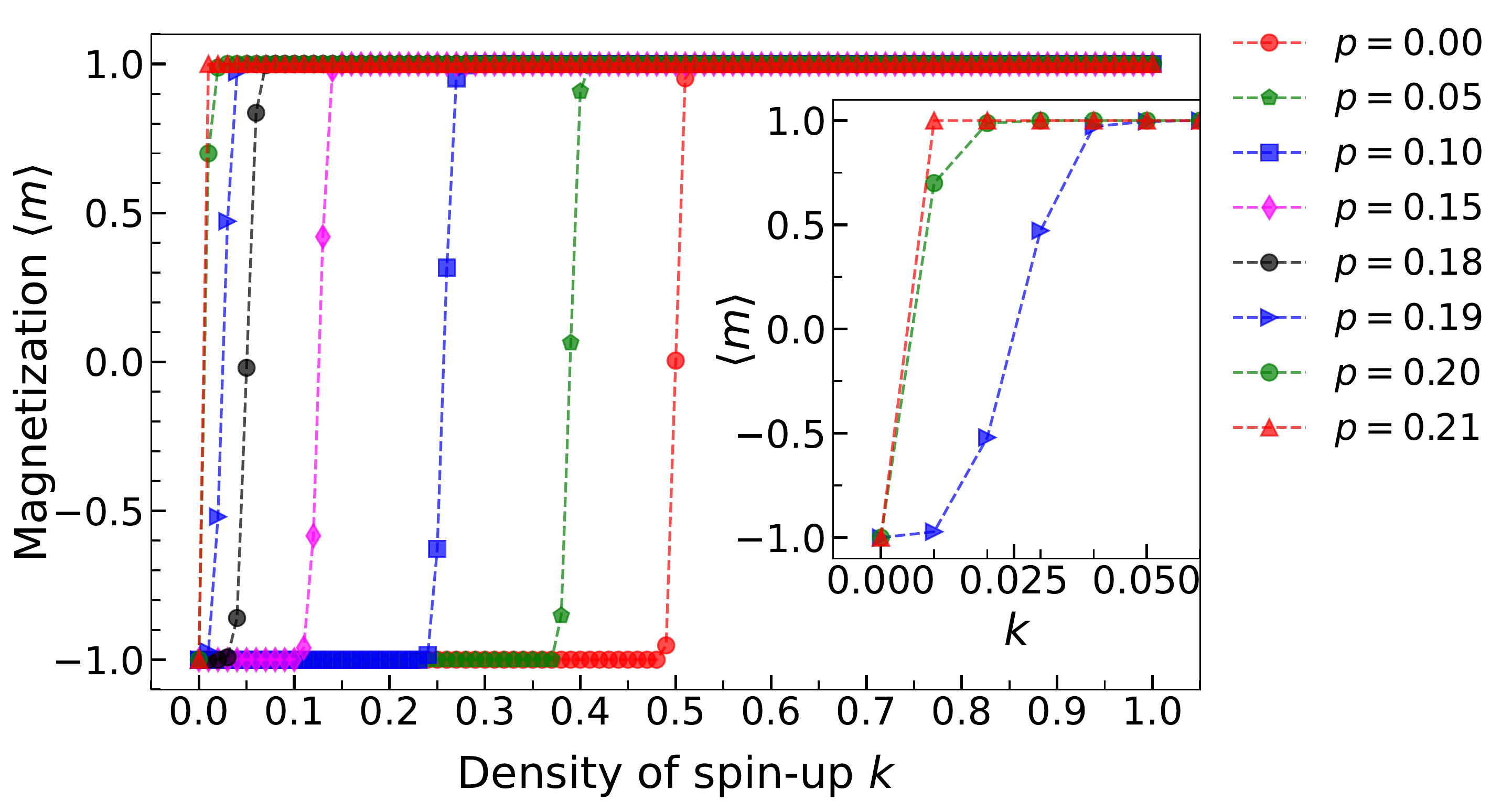}
    \caption{The numerical result of the order parameter $m$ versus density of spin-up $k$ for typical of the probability of the mass media effect $p$ based on the $q$-voter model on the complete graph. As seen, the probability $p$ eliminates the phase transition at $p \geq  0.21$. In other words, the phase transition is absent for $p \geq 0.21$; at a small density of spin-up $k = 0.01$, the system reaches a homogeneous state with all agents at the end having the same opinion ($m = 1$). The inset graph displays in more detail for $p = 0.19, 0.20,$ and $p = 0.21$. Each data point averages $500$ samples, and the population size $N = 10000$. }
    \label{fig:fraction_q}
\end{figure}

In the $q$-voter model case, we find that the effect of the probability $p$ eliminates the usual phase transition at the probability $ p \approx  0.21$, namely, for a small of initial spin-up density $k = 0.01$ the system reaches a homogeneous state with $m = + 1$ at $p \approx 0.21$ as exhibited in Fig.~\ref{fig:fraction_q}. In other words, the usual phase transition exists for the probability $p <  0.21$, i.e. we will obtain the magnetization always less than $+1$ for $p <  0.21$. In addition, we will find that the system always reaches an homogeneous state with $m = +1 $ for $p \geq 0.21$. The inset graph of Fig.~\ref{fig:fraction_q} displays in more detail for the range $k = 0.00 - 0.05$ and obtains, at a small $k = 0.01$ the magnetization $m$ for typical values of $p = 0.20\, (m \approx 0.70), p = 0.21\, (m \approx 1.0)$.

\textcolor{black}{The critical point $d_c$ of the $q$-voter model for the probability $p = 0.1$ is obtained at the spin-up density $k = k_c \approx 0.256$ with the same scaling parameters as those obtained in the majority rule model, namely $\delta_1 \approx 0 $ and $\delta_2 \approx 0.5$. In this case, the scaling parameters work for $p < p_t \approx 0.21$, and for $p > p_t$ we find the scaling parameters which make the best collapse of all data are $\delta_1 \approx 0$ and $\delta_2 > 0.5$. For example, for the probabilities of the external field $p = 0.25$ and $p = 0.5$, we find the scaling parameters $\delta_2 \approx 0.82$ and $\delta_2 \approx 1.0$, respectively (graph not shown). This means that, the scaling parameters $\delta_1 \approx 0$ and $\delta_2 \approx 0.5$ are not universal for all values of $p$, as in the majority rule model case. However, this paper does not discuss the exact limit value of the scaling parameters $\delta_1$ and $\delta_2$ to the external field $p$.}

\begin{figure}[tb]
    \centering
    \includegraphics[width = 10 cm ]{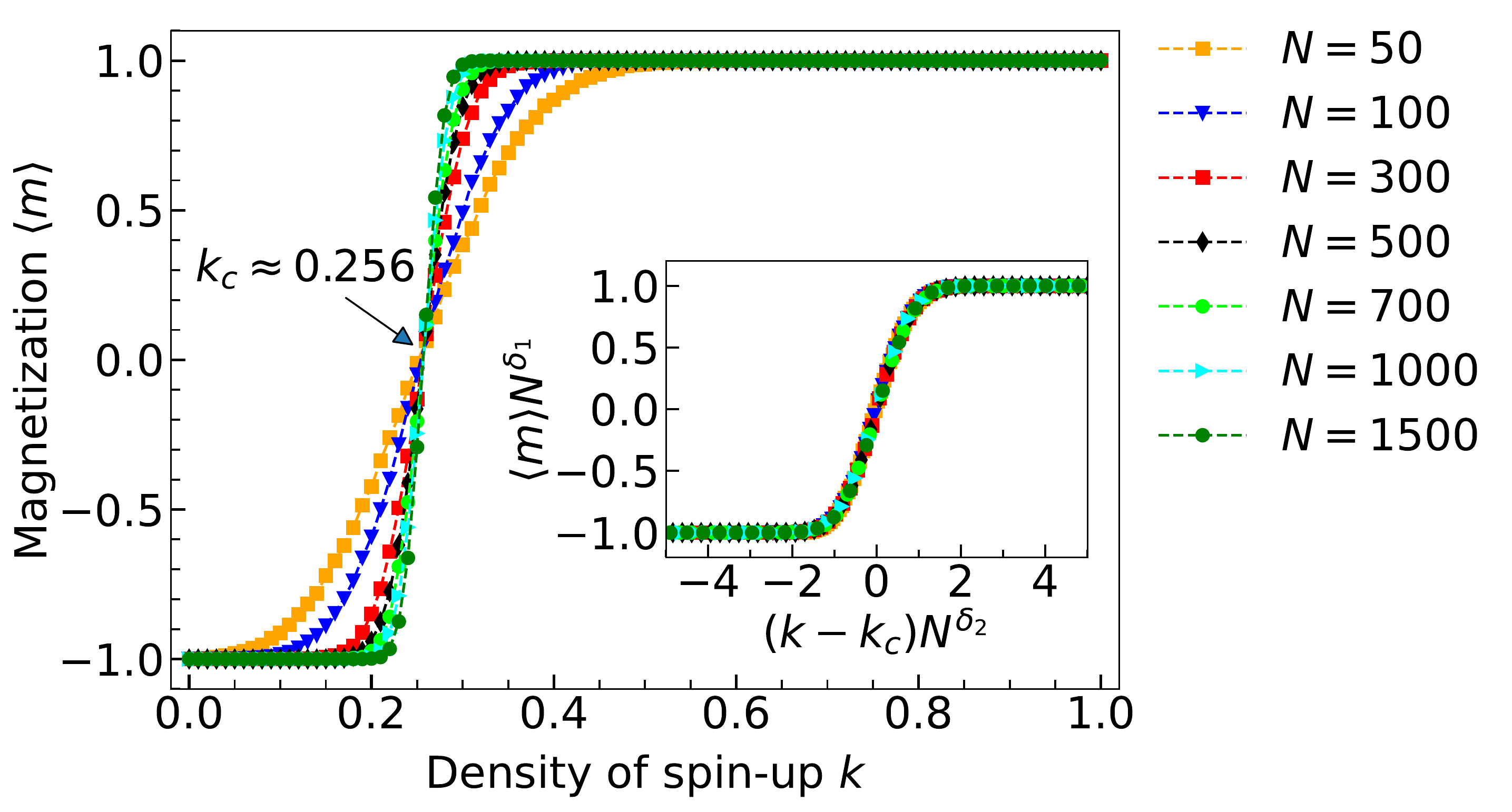}
    \caption{{\color{black}(Main graph) Numerical result of the average magnetization $m$ versus the density spin-up $k$ of the $q$-voter model for the probability mass media $p = 0.1$ and typical of the population size $N$. The critical point {\color{black}$ k_c \approx 0.256$} is obtained from crossing lines $m$. The inset graph displays the scaling plot of $m$. Using finite-size scaling analysis, the scaling parameters that make the best collapse of all data are $\delta_1 \approx 0, \delta_2 \approx 0.5$.}}
    \label{fig:variate_N_qvoter}
\end{figure}

\begin{figure}[tb]
    \centering
    \includegraphics[width = 0.9\linewidth]{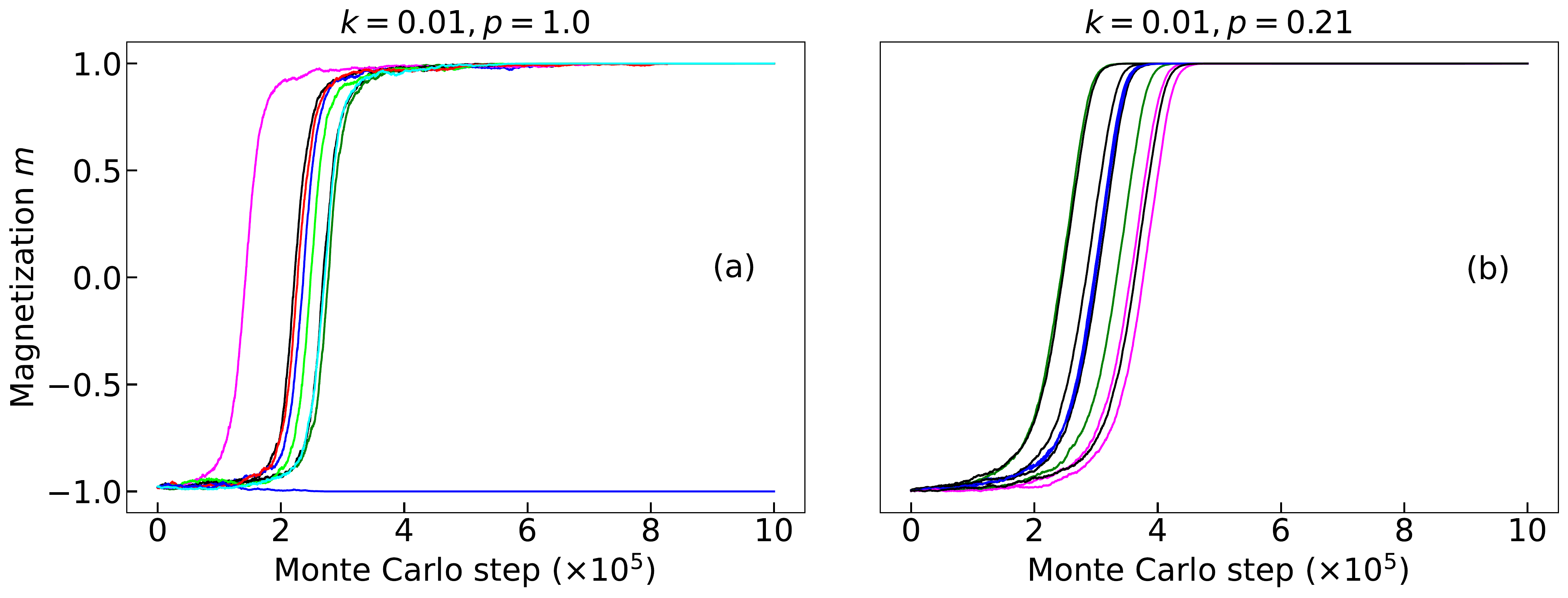}
    \caption{Evolution of the magnetization $m$ under the influence of mass media based on the majority rule (a) and the $q$-voter models (b) on the complete graph. (a) As seen, for a small density of spin-up $k = 0.01$ and the probability mass media effect $p = 1.0$ there is a data reaches $m = -1$ indicates that the mass media effect not strong to make the system reaches a homogeneous state (complete consensus) with $m = 1$. In other words, the phase transition is observed in this model. (b) In the $q$-voter model case, for a small density of spin-up $k =0.01$ and the probability of mass media $p = 0.21$, all samples reach a homogeneous state with $m =  1$, indicate that the the mass media effect eliminates the phase transition at $p \geq 0.21$. The system size $N = 10000$.}
    \label{fig:mcsp1}
\end{figure}

We check the evolution of the magnetization $m$ in time (Monte Carlo step) for a small density of spin-up $k = 0.01$ and the population size $N = 10000$ for the maximum of the external field effect $p = 1$ for the majority model case and $p = 0.21$ for the $q$-voter model case  as exhibited in Fig.~\eqref{fig:mcsp1}. One can see that, in the majority rule model case, there is a sample that shows the system reaching a homogeneous state with $m = -1$ even for $p =  1$, indicating that the phase transition is observed in this model for all values of $p$, as exhibited in Fig.~\eqref{fig:mcsp1} (a). In $q$-voter model case as exhibited in Fig.~\eqref{fig:mcsp1} (b), for a small $k = 0.01$ and $p = 0.21$ all samples reach a homogeneous state with $m = +1$ indicates the absence of the phase phase transition for $p \geq 0.21$ in this model.
 
\textcolor{black}{If we consider the scaling parameters of both models, we find the same scaling parameters, namely $\delta_1 \approx 0$ and $\delta_2 \approx 0.5$. The same scaling parameter means that both models are similar in terms of universality class because both models are defined on the same graph, namely a complete graph, even though the two models have different microscopic interactions. However, it is not necessary that if the models are defined on the same graph, they will have the same universality class \cite{muslim4241509phase}. The value of this scaling parameter depends on the model's topology, especially the model's network structure. For example, the scaling parameters of the model defined on the complete graph will be different from the model defined on the two-dimensional square lattice \cite{muslim4241509phase}}

\begin{figure}[t!]
    \centering
    \includegraphics[width = 0.9\textwidth]{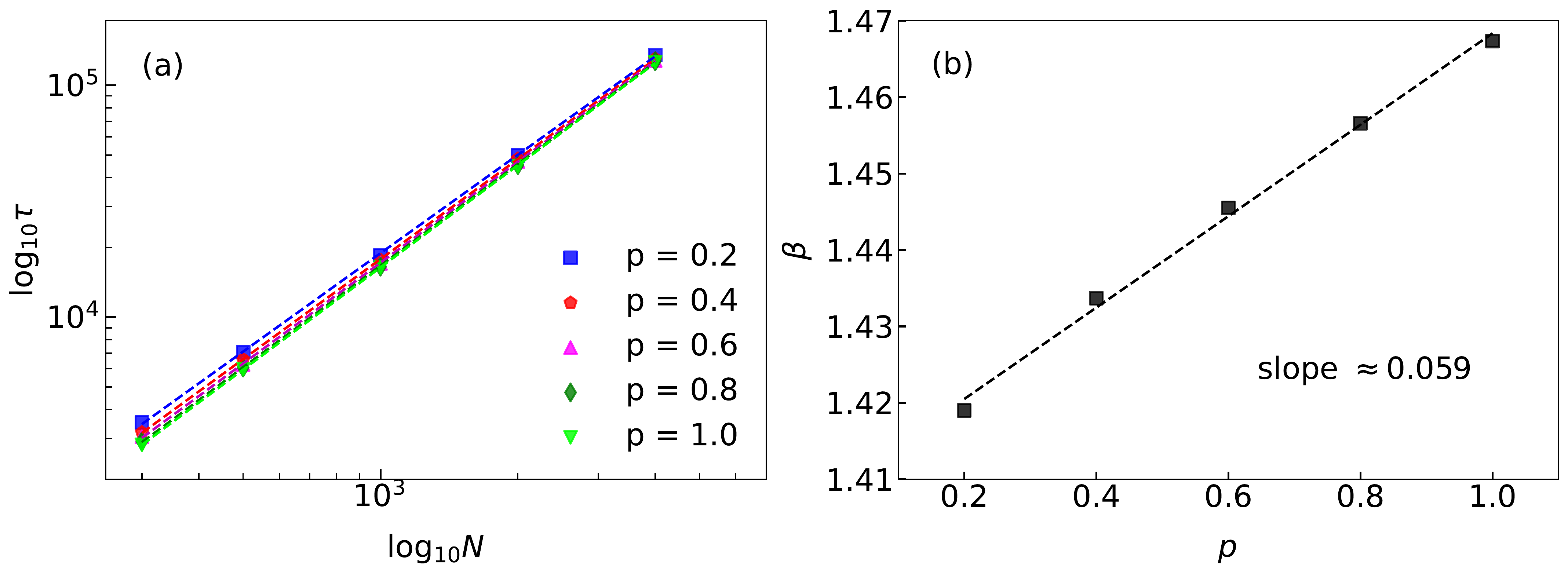}
    \caption{ \textcolor{black}{(a) Log-log plot of the relaxation time $\tau$ versus population size $N$ of the majority rule model for typical values $p$.  The model follows the power-law relation $\tau \sim N^{\beta}$, where $\beta$ is the slope of the data, depending on the probability $p$. The density of spin-up $k = 0.5$ and each data averages of $10000$ samples. The effect of the probability $p$ reduces the relaxation time $\tau$. (b) Normal plot of the slope $\beta$ versus the probability $p$, where $\beta$ follows the linear relation $ \beta \sim \lambda \,p$, $\lambda \approx 0.059$ is the best fit of the data.}}
    \label{fig:relax}
\end{figure}

The interesting quantity in the Monte Carlo simulation is the relaxation time $\tau$, which is the time needed to reach a homogeneous state (all agents at the end have the same opinion). In this part, we want to find the relation of the relaxation time $\tau$ to the system size $N$. We variate $N = 300, 500, 1000, 2000$ and $4000$ and each of the data point averages of $10000$ samples. The log-log plot of the relaxation time $\tau$  versus the probability  $p$ of the majority rule model is exhibited in Fig.~\ref{fig:relax}~(a). One can see that the external field's effect decreases the relaxation time as $p$ increases. The model follows the power-law relation as follows:
\begin{equation}\label{eq:relax_log}
    \tau \sim N^{\beta},
\end{equation}
where $\beta$ is a slope of the data and depends on the probability $p$. The dependence of $\beta$ to the probability $p$ satisfies a linear relation as $\beta \sim \lambda \,p$, where $\lambda \approx 0.059$ is the best fitting of the data as exhibited in the inset of Fig.~\ref{fig:relax}~(b). The power-law relation such in Eq.~\eqref{eq:relax_log} is also obtained in the other opinion dynamics models with various scenarios \cite{sousa2008effects, crokidakis2012effects}.

\begin{figure}[t!]
    \centering
    \includegraphics[width = 0.9\textwidth]{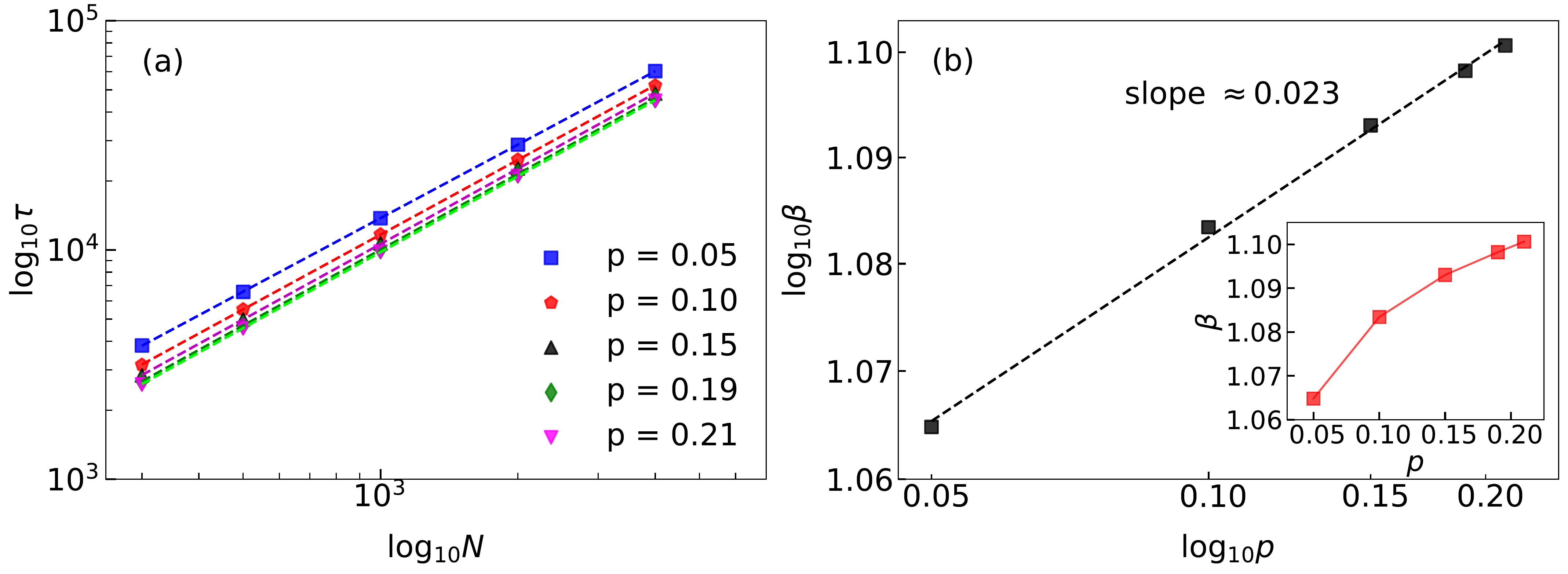}
    \caption{\textcolor{black}{(a) Log-log plot of the relaxation time $\tau$ versus population size $N$ of the $q$-voter model for typical values $p$.  The model follows the power-law relation $\tau \sim N^{\beta}$, where $\beta$ is the slope of the data, depending on the probability $p$. The density of spin-up $k = 0.5$ and each data averages of $10000$ samples. It can be seen that the probability $p$ reduces the relaxation time $\tau$ as $p$ increases. (b) (main graph) Log-log plot of the slope $\beta$ versus the probability $p$, where $\beta$ follows the power-law relation $\beta \sim p^{\lambda}$, $\lambda \approx 0.023$ is the best fit of the data.}}
    \label{fig:relax_q}
\end{figure}

In the $q$-voter model case, we also variate $N = 300, 500, 1000, 2000$ and $4000$ and each data averages of $10000$ samples. In Fig.~\ref{fig:relax_q}~(a), the effect of the external field also decreases the relaxation time $\tau$ as $p$ increases as in the majority model case. The model also follows the power-law relation in \eqref{eq:relax_log}, where $\beta$ depends on the probability $p$, which follows a power-law relation $\beta \sim p^{\lambda}$, where $\lambda \approx 0.023$ is the best slope of the data as exhibited in Fig.~\ref{fig:relax_q}~(b).

{\color{black} The dependence of the spin-up density $k$ to the time evolution $t$ or the relaxation time $\tau$ can be evaluated from the master equation that can be expressed as  \cite{krapivsky2010kinetic}:
\begin{equation}\label{eq:rateofd}
    \dfrac{dk}{dt}= P_{+}-P_{-},
\end{equation}
where $P_+$ and $P_-$ are the probabilities of spin-up increases and decreases, respectively, and depend on the model. For example, the probabilities of spin-up $P_+$ and $P_-$ for the $q$-voter model (with $q=5$) with the external field $p$ for the finite system can be expressed as follows:

\begin{equation}\label{eq8}
    \begin{aligned}
        & P_{+} = \dfrac{\prod_{i=0}^{4} \left(N_u-i\right)N_d}{\prod_{i = 0}^{5} \left(N-i\right)}+ \dfrac{p\,N_{d}}{N} \left[1- \dfrac{ \prod_{i=1}^{4} \left(N_{u}-i\right)-\prod_{i=0}^{4} \left(N_{d}-i\right)  }{\prod_{i=0}^{4}\left(N-i\right)}\right]\\
        & P_{-} = \dfrac{\prod_{i=0}^{4} \left(N_d-i\right)N_u}{\prod_{i = 0}^{5} \left(N-i\right)},
    \end{aligned}
\end{equation}
where $N_u$ and $N_d$ are the total spin-up and spin-down, respectively. For the infinite system Eq.~\eqref{eq8} reduce to $P_{+} =  k^5\left(1-k\right) + p\left(1-k\right)\left[ 1-k^5-(1-k)^5\right]$ and $P_{-} = k \left(1-k\right)^5$, where $k = N_{u}/N$ is the spin-up density. The general solution of $k$ in Eq.~\eqref{eq:rateofd} respects to the relaxation time $\tau$  can be obtained as follows:
\begin{equation}
    \tau = \int_{k_{i}}^{k_f} \dfrac{dk}{P_{+}-P_{-}},
\end{equation}
where $k_i$ and $k_f$ are the initial and final of the spin-up density related to the initial time ($t_i = 0$) and the final time (relaxation time) $\tau$, respectively.} \textcolor{black}{The explicit expression of spin-up density  $k$ to the relaxation time $\tau$ is not discussed in this paper. However, the dependence of the spin-up density $k$ respect to the relaxation time $\tau$ in general is in the exponential form, $k \sim \exp{\pm \tau}$, and depends on $\beta$ in the exponential form just by inserting Eq.~\eqref{eq:relax_log} to the general solution of density spin-up $k$.}

Finally, our numerical results suggest several similarities and differences due to the external field effect of both models in this paper. For the similarities, (1) based on the similarity of the scaling parameter, both models are identical, (2) the relaxation time $\tau$ of both models follows the power-law relation in Eq.~\ref{eq:relax_log}, and the external field effect decreases the relaxation time $\tau$ as $p$ increases as shown in Figs.~\ref{fig:relax}~(a) and \ref{fig:relax_q}~(a). For the differences, the effect of the external field does not eliminate the usual phase transition in the majority rule model. However, in the $q$-voter model, the external field eliminates the usual phase transition at $p >\sim 0.21$. In addition, the slope of the relation relaxation time $\tau$ and the population size $N$ have different relation forms. In the majority rule model, the slope $\beta$ follows a linear relation $\beta \sim 0.059 \,p$ and in the $q$-voter model follows the power-law relation $\beta \sim p^{0.023}$. In addition, based on the sociophysics point of view, our result suggests that the effect of the external field (mass media) in society can lead the opinion reaches a complete consensus (depending on the scenario), although the initial spin-up is very small (initial of minority opinion).

\section{Final Remark}
This paper investigates the external field effect (mass media) on opinion formation based on the majority rule and the $q$-voter models on the complete graph. We consider a small group consists six agents to follow the majority rule and the $q$-voter models. In the majority rule model, two agents follow the opinion of four agents when there is a majority opinion or consensus. Otherwise, two agents follow the mass media opinion with probability $p$. In the $q$-voter model, the voter adopts the $q$-sized agents if the  $q$-sized agents have the same opinion (homogeneous). Otherwise, the voter follows the mass media opinion with probability $p$.

We perform Monte Carlo simulation for both models to analyze the external field effect on the system. Based on our result, the existence of the external field in the majority model is not strong enough to make the system reaches a homogeneous state with magnetization $m = +1$ even for a maximum $p =  1$, which indicates the existence of the usual phase transition for all values of $p = \{0,1\}$. In the $q$-voter model, the probability $p$ eliminates the usual phase transition at $p \approx 0.21$ which indicating the existence of the usual phase transition for $p < 0.21$. We will find that the system will always reach a homogeneous with $p \geq 0.21$ even for a small density spin-up $k = 0.01$.

We find the critical point and the scaling parameters of both models using finite-size scaling relations in \eqref{eq4}-\eqref{eq5}. We find that the critical point $d_c$ of the majority rule and $q$-voter models are $d_c \approx 0.379$ for $p = 0.2$ and $d_c \approx 0.256$ for $p = 0.1$, respectively. The scaling parameters related to the magnetization $m$ of both models are the same, namely $\delta_1 \approx 0$, and $\delta_2 \approx 0.5$. These parameters are universal for all values of population size $N$ and the probability of the external field $p$ in the phase transition range.

We also analyze the relaxation time of both models $\tau$, namely the time needed for the system to reach a homogeneous state. Our result suggests that the external field effect decreases the relaxation time of both models. The relaxation time of both models follow the power-law relation $\tau \sim N^{\beta}$, where $\beta$ depends on the probability $p$. In the majority rule model case, $\beta$ follows the linear relation $\beta \sim \lambda\,p$, where $\lambda \approx 0.059$ (see Fig.~\eqref{fig:relax}~(b)) and in the $q$-voter model case, $\beta$ follows the power-law relation $\beta \sim p^{\lambda}$, where $\lambda \approx 0.023$ (see Fig.~\ref{fig:relax_q}~(b)). Such power-law relation is observed in other opinion dynamics models, such as the two-dimensional Sznajd model with the mass media effect \cite{crokidakis2012effects}.


\section*{Declaration of Interests}
The authors declare that they have no known competing financial interests or personal relationships that could have appeared to influence the work reported in this paper.

\section*{Acknowledgments}
The authors would like to thank Universitas Sumatra Utara for its financial support and BRIN Research Center for Quantum Physics for providing access to its mini-HPC to perform numerical simulations.



\bibliographystyle{elsarticle-num} 
\bibliography{cas-refs}

\end{document}